\documentclass[reprint,amsmath,amssymb,aps,prb,letterpaper, superscriptaddress]{revtex4-1}

\usepackage{graphicx}% Include figure files
\usepackage{epstopdf}
\usepackage{amsmath}
\usepackage{dcolumn}% Align table columns on decimal point
\usepackage{bm}% bold math
\usepackage{mathtools}
\usepackage[mathlines]{lineno}% Enable numbering of text and display math
\begin{document}

\preprint{APS/123-QED}

\newcommand{\UP} {URu$_{2}$Si$_{2}$ }
\newcommand{\UFe} {URu$_{2-x}$Fe$_{x}$Si$_{2}$ }

\title{Phase diagram and thermal expansion measurements on the system of URu$_{2-x}$Fe$_{x}$Si$_{2}$}% Force line breaks with \\

\author{S.~Ran}
\affiliation{Department of Physics, University of California, San Diego, USA}
\affiliation{Center for Advanced Nanoscience, University of California, San Diego, USA}
\author{C.~T.~Wolowiec}
\affiliation{Department of Physics, University of California, San Diego, USA}
\affiliation{Center for Advanced Nanoscience, University of California, San Diego, USA}
\author{I.~Jeon}
\affiliation{Department of Physics, University of California, San Diego, USA}
\affiliation{Center for Advanced Nanoscience, University of California, San Diego, USA}
\affiliation{Materials science and Engineering program, University of California, San Diego, USA}
\author{N.~Pouse}
\affiliation{Department of Physics, University of California, San Diego, USA}
\affiliation{Center for Advanced Nanoscience, University of California, San Diego, USA}
\author{N.~Kanchanavatee}
\affiliation{Department of Physics, University of California, San Diego, USA}
\affiliation{Center for Advanced Nanoscience, University of California, San Diego, USA}
\author{K.~Huang}
\affiliation{Department of Physics, University of California, San Diego, USA}
\affiliation{Center for Advanced Nanoscience, University of California, San Diego, USA}
\affiliation{Materials science and Engineering program, University of California, San Diego, USA}
\author{D. Martien}
\affiliation{Quantum Design, Inc., San Diego, USA}
\author{T. DaPron}
\affiliation{Quantum Design, Inc., San Diego, USA}
\author{D. Snow}
\affiliation{Quantum Design, Inc., San Diego, USA}
\author{M. Williamsen}
\affiliation{Quantum Design, Inc., San Diego, USA}
\author{S. Spagna}
\affiliation{Quantum Design, Inc., San Diego, USA}
\author{M. B. Maple}
\affiliation{Department of Physics, University of California, San Diego, USA}
\affiliation{Center for Advanced Nanoscience, University of California, San Diego, USA}
\affiliation{Materials science and Engineering program, University of California, San Diego, USA}

\date{\today}% It is always \today, today,

\begin{abstract}
Thermal expansion, electrical resistivity, magnetization, and specific heat measurements were performed on URu$_{2-x}$Fe$_{x}$Si$_{2}$ single crystals for various values of the Fe concentration {\itshape x} in both the hidden order (HO) and large moment antiferromagnetic (LMAFM) regions of the phase diagram. Our results show that the paramagnetic (PM) to HO and LMAFM phase transitions are manifested differently in the thermal expansion coefficient.  For Fe concentrations near the boundary between the HO and LMAFM phases at $x_c$ $\approx$ 0.1, we observe two features in the thermal expansion upon cooling, one that appears to be associated with the transition from the PM to the HO phase and another one at lower temperature that may be due to the transition from the HO to the LMAFM phase.  These two features have not been observed in other measurements such as specific heat or neutron scattering. In addition, the uniaxial pressure derivative of the transition temperature, based on a calculation using thermal expansion and specific heat data, changes dramatically when crossing from the HO to the LMAFM phase.    

\end{abstract}


\maketitle

%\tableofcontents

\section{\label{sec:level1}Introduction}

The search for the order parameter of the hidden order (HO) phase in \UP has attracted an enormous amount of attention for the past three decades.~\cite{Palstra85,Maple86,Schlabitz86,Mydosh11} The small antiferromagnetic moment of only $\sim$0.03 $\mu_{B}$/U found in the HO phase is too small to account for the entropy of $\sim$ 0.2Rln(2) derived from the second-order mean field BCS-like specific heat anomaly associated with the HO transition that occurs below {\itshape T}$_{0}$ = 17.5~K.~\cite{Maple86,Broholm87} A first-order transition from the HO phase to a large moment antiferromagnetic (LMAFM) phase occurs under pressure at a critical pressure {\itshape P}$_{c}$ that lies in the range 0.5 - 1.5~GPa.~\cite{Amitsuka99,Jeffries07,Motoyama08,Butch10} Many studies suggest that the HO and LMAFM phases are intimately related and that a comprehensive investigation of both phases will be useful in unraveling the nature of the order parameter of the HO phase.~\cite{Bourdarot10} Although the order parameters are presumably different in the HO and LMAFM phases, the two phases exhibit almost indistinguishable transport and thermodynamic properties. This behavior has been referred to as \textquotedblleft adiabatic continuity$\textquotedblright$.~\cite{Jo98}

We have recently demonstrated that tuning \UP by substitution of Fe for Ru affords the opportunity to study both the HO and LMAFM phases and the HO-LMAFM transition at atmospheric pressure.~\cite{Kanchanavatee11} Specifically, the substitution of the smaller Fe ions for Ru ions in \UP appears to act as a chemical pressure such that the temperature vs.~Fe substitution phase diagram for the URu$_{2-x}$Fe$_{x}$Si$_{2}$ system resembles the temperature vs.~applied pressure phase diagram for URu$_{2}$Si$_{2}$. In a previous study, neutron diffraction measurements were carried out on single crystal samples of \UFe for various values of {\itshape x}.~\cite{Das15} The results revealed that the magnetic moment increases abruptly to a maximum value at {\itshape x} = 0.2, above which it then decreases slowly with {\itshape x}, supporting the interpretation that tuning by Fe substitution acts as a chemical pressure. Therefore, we have a unique opportunity to perform experiments at ambient pressure that will allow us to study, with confidence, the LMAFM phase that is known to exist under high pressure.

On the other hand, the phase boundary between the HO and LMAFM phases has not been definitively determined for the URu$_{2-x}$Fe$_{x}$Si$_{2}$ system. Extensive effort has been expended to map out the precise phase boundary between the HO and LMAFM phases in \UP under pressure, which is not vertical on the {\itshape T}-{\itshape P} phase diagram.~\cite{Butch10,Motoyama08,Bourdarot05,Amitsuka07,Amitsuka08,Hassinger08,Niklowitz10} The critical pressure at which the HO-LMAFM transition occurs is about 1.5~GPa at {\itshape T}$_{0}$, while it drops to about 0.8~GPa at the base temperature.~\cite{Butch10} Therefore, at intermediate values of pressure, e.g., 1~GPa, \UP goes through two successive phase transitions upon cooling: a second order transition from the paramagnetic (PM) into the HO phase, and then a first order transition from the HO into the LMAFM phase. This has not been observed in either polycrystalline or single crystal samples of URu$_{2-x}$Fe$_{x}$Si$_{2}$. It is possible that in the polycrystalline samples, both the PM-HO and HO-LMAFM transitions are broadened, especially in the vicinity of the HO-LMAFM phase boundary, so that the transition from the HO phase into the LMAFM phase with decreasing temperature is not readily discernible. Neutron diffraction measurements have been carried out on the single crystals, but only for a few selected concentrations.~\cite{Das15} 

In the experiments reported herein, we performed thermal expansion, electrical resistivity, magnetization, and specific heat measurements on URu$_{2-x}$Fe$_{x}$Si$_{2}$ single crystals for various values of {\itshape x} throughout both the HO and LMAFM regions of the phase diagram.  Interestingly, the transition from the HO to the LMAFM phase with decreasing temperature in \UP that occurs under pressure was 
previously observed by means of thermal expansion measurements.~\cite{Motoyama03,Kambe13}  Our thermal expansion measurements reveal differences in the features associated with the HO and LMAFM phase transitions that appear in the thermal expansion coefficient, reflecting differences in the coupling of the two phases to the lattice.  For Fe concentrations near the boundary between the HO and LMAFM phases at $x_c$ $\approx$ 0.1, two features in the thermal expansion are found upon cooling, one that appears to be associated with the transition from the PM to the HO phase and another at lower temperature that may be due to the transition from the HO to the LMAFM phase.  These two features have not been observed  in other measurements such as specific heat or neutron scattering.  In addition, the uniaxial pressure derivative of the transition temperature, based on a calculation using thermal expansion and specific heat data, changes dramatically when crossing from the HO to the LMAFM phase. 

\section{\label{sec:level1}Experimental methods}

Single crystals of Fe substituted \UP were grown by the Czochralski method in a tetra-arc furnace. Electrical resistivity measurements were performed using a home-built probe in a liquid $^{4}$He Dewar by means of a standard four-wire technique at 16 Hz using a Linear Research LR700 ac resistance bridge. Magnetization measurements were made in magnetic fields of 0.1 T using a Quantum Design magnetic property measurement system (MPMS). Specific heat measurements were performed in a Quantum Design Dynacool Physical Property Measurement System (PPMS) using a heat-pulse technique. Thermal expansion measurements were made in a Quantum Design PPMS with a dilatommetry option.

\section{Results}
\begin{figure}[!htbp]
\begin{center}
\includegraphics[angle=0,width=80mm]{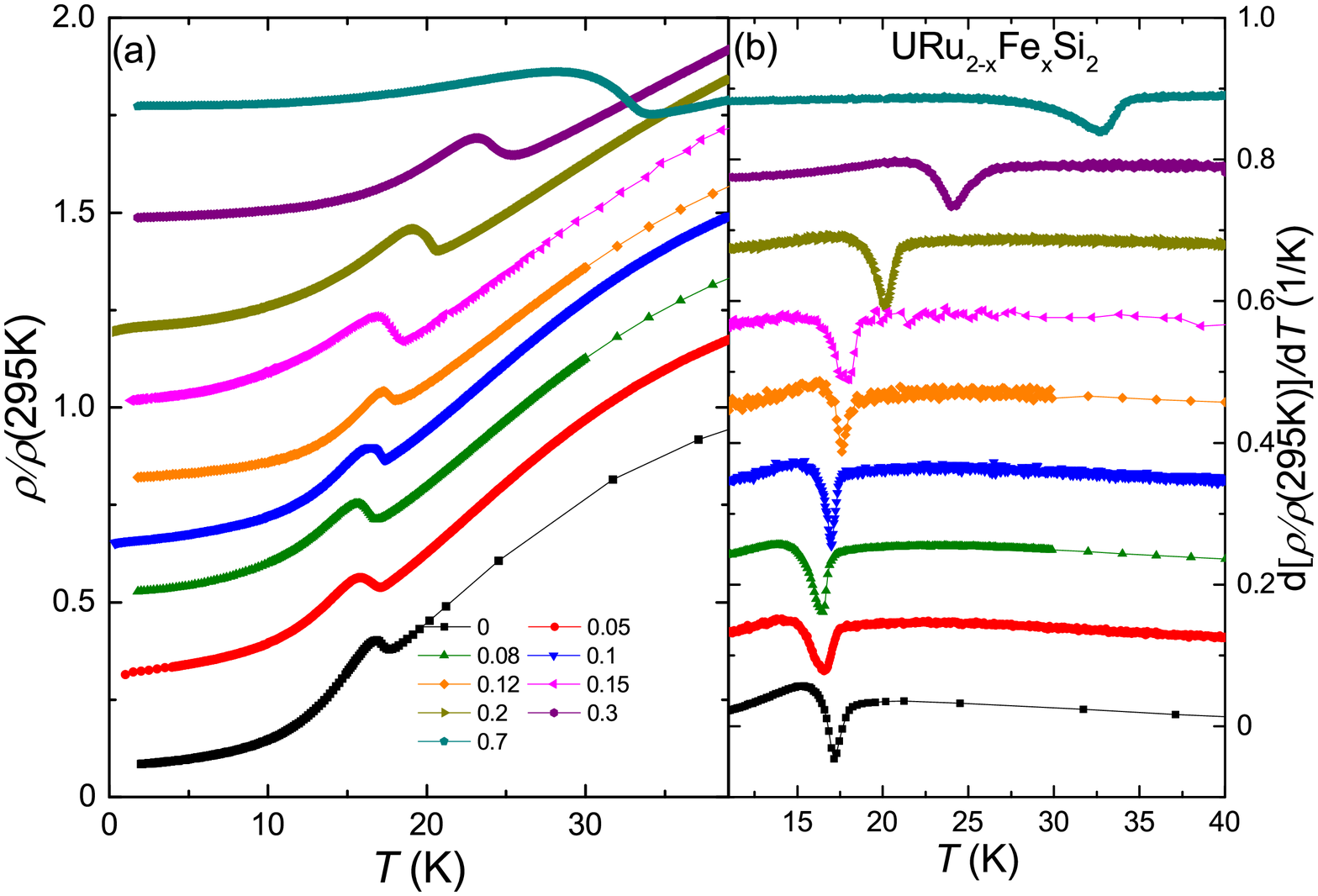}
\end{center}
\caption{(Color online) (a) Electrical resistivity $\rho$ and (b) derivative of $\rho$, d$\rho$/d{\itshape T}, vs.~temperature {\itshape T} for \UFe single crystals with various values of {\itshape x} between 0 and 0.7. Both quantities are normalized to the room temperature values.}
\label{RT}
\end{figure}

\begin{figure}[!htbp]
\begin{center}
\includegraphics[angle=0,width=80mm]{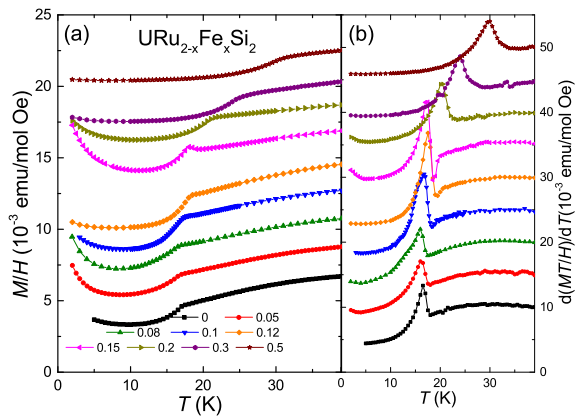}
\end{center}
\caption{(Color online) (a) Magnetization $M$ and (b) derivative of $MT$, d({\itshape MT})/d{\itshape T}, vs.~temperature {\itshape T} for \UFe single crystals with various values of {\itshape x} between 0 and 0.7.}
\label{MT}
\end{figure}

Shown in Figure~\ref{RT}(a) are electrical resistivity $\rho(T)$ data, normalized to room temperature values, for various URu$_{2-x}$Fe$_{x}$Si$_{2}$ compounds. The $\rho(T)$ curves are offset vertically for clarity. For this study, we focus on the interrelation of the HO and LMAFM phases. Therefore, we did not perform low temperature measurements to study superconductivity. The transition from the PM into the HO phase in the parent compound is manifested as an anomaly at around 17~K. The transition temperature can be extracted from the minimum in d$\rho$/d{\itshape T}. Upon Fe substitution, the signature of the phase transition is preserved, while the transition temperature {\itshape T}$_{0}$ changes systematically. After an initial suppression down to 16.2~K at {\itshape x} = 0.08, {\itshape T}$_{0}$ increases up to 34~K at {\itshape x} = 0.7. Similar results are obtained from magnetization {\itshape M}({\itshape T}) data, as shown in Fig.~\ref{MT}. The corresponding feature for the phase transition in {\itshape M}({\itshape T}) is the slope change. This can be seen more clearly from the quantity d({\itshape MT})/d{\itshape T}, which is expected to yield a feature that is similar in shape to that observed in the specific heat data.~\cite{Fisher62}  Although the signature of the phase transition in both $\rho$({\itshape T}) and {\itshape M}({\itshape T}) seems to remain unchanged throughout the entire Fe concentration range sampled in this study, the neutron diffraction experiments indicate that the ground state of \UFe changes from the HO to the LMAFM phase, with a phase boundary close to {\itshape x}$_c$ = 0.1.~\cite{Das15}

\begin{figure}[!htbp]
\begin{center}
\includegraphics[angle=0,width=80mm]{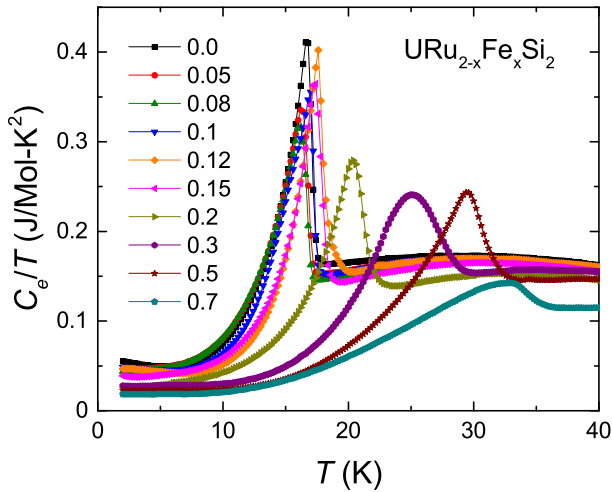}
\end{center}
\caption{(Color online) Electronic specific heat divided by temperature, {\itshape C}$_{e}$/{\itshape T}, vs.~temperature {\itshape T} for \UFe single crystals with various values of {\itshape x} between 0 and 0.7.}
\label{Cp}
\end{figure}

Displayed in Figure~\ref{Cp} is the electronic contribution to the specific heat, {\itshape C}$_{e}$({\itshape T}), divided by temperature {\itshape T}, vs $T$, determined by subtracting the phonon contribution to the specific heat {\itshape C}$_{ph}$({\itshape T}) from the measured specific heat {\itshape C}({\itshape T}), as discussed in our previous work.~\cite{Kanchanavatee11} The electronic specific heat {\itshape C}$_{e}$({\itshape T}) exhibits a well-defined BCS-like anomaly upon transition from the PM into the HO and LMAFM phases at {\itshape T}$_{0}$. The size of the jump at the transition decreases above {\itshape x} = 0.2 and broadens considerably at {\itshape x} = 0.7. In our previous work on polycrystalline samples, a broad shoulder above {\itshape T}$_{0}$ was observed and attributed to disorder~\cite{Kanchanavatee11}. There is no sign of the broad shoulder in the {\itshape C}$_{e}$({\itshape T}) vs. {\itshape T} data for the single crystals presented here, showing that the single crystals do not suffer from similar problems.

\begin{figure}[!htbp]
\begin{center}
\includegraphics[angle=0,width=80mm]{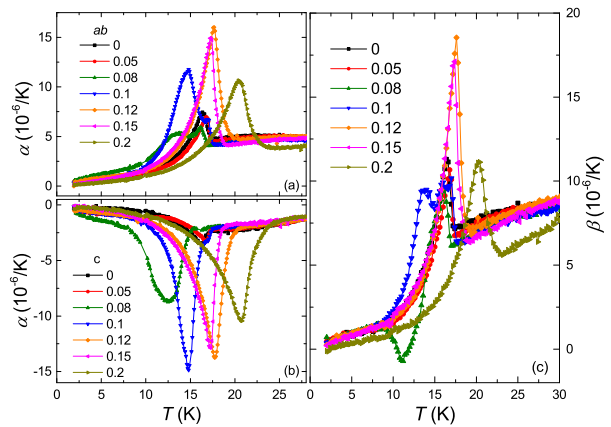}
\end{center}
\caption{(Color online) Linear thermal expansion coefficients in (a) the {\itshape ab}-plane ($\alpha_{ab}$) and (b) along the {\itshape c}-axis ($\alpha_{c}$) vs. temperature {\itshape T} for \UFe single crystals with various values of {\itshape x} between 0 and 0.2.  (c) The calculated volume thermal expansion coefficient $\beta$, derived from the $\alpha_{ab}$ and $\alpha_{c}$ vs. {\itshape T} data in (a) and (b), respectively.}

%Temperature dependent linear thermal expansion coefficient data for \UFe compounds for various values of {\itshape x} between 0 and 0.2 in (a) the {\itshape ab}-plane ($\alpha_{ab}$) and (b) along the {\itshape c}-axis ($\alpha_{c}$). Data for the calculated volume thermal expansion coefficient based on $\alpha_{ab}$ and $\alpha_{c}$, are shown in (c).}
\label{TE}
\end{figure}

Presented in Figures~\ref{TE}(a) and (b) are the linear thermal expansion coefficients in the {\itshape ab}-plane, $\alpha_{ab}$, and along the {\itshape c}-axis, $\alpha_{c}$, vs.~temperarture {\itshape T} for \UFe single crystals with various values of {\itshape x} between 0 and 0.7. The linear thermal expansion coefficient is strongly anisotropic, with $\alpha_{ab}$ positive and $\alpha_{c}$ negative. At the HO-LMAFM phase transition, an anomaly is observed in both $\alpha_{ab}$ and $\alpha_{c}$. However, the signature of the anomaly is markedly different for the PM to HO and LMAFM transitions, showing that the HO and LMAFM phases are clearly distinct from one another. For {\itshape x} $\textless$ 0.05, where the compounds exhibit a PM-HO phase transition, the size of the jump at {\itshape T}$_{0}$ is relatively weak, while, for larger {\itshape x}, where the LMAFM phase is the ground state, the size of the jump at {\itshape T}$_{0}$ is more than three times larger. Similar differences in the size of the jumps in 
$\alpha_{ab}$ and $\alpha_{c}$ for the PM-HO and PM-LMAFM transitions have been reported for \UP under pressure.~\cite{Motoyama03} One of the basic features of \UP is the significant amount of coupling of the HO phase to the lattice.~\cite{Visser86} However, the net volume change is even larger for the PM-LMAFM phase transition, as shown in Fig.~\ref{TE}(c), indicating that the LMAFM phase is even more strongly coupled to the lattice than the HO phase. 

\begin{figure}[!htbp]
\begin{center}
\includegraphics[angle=0,width=80mm]{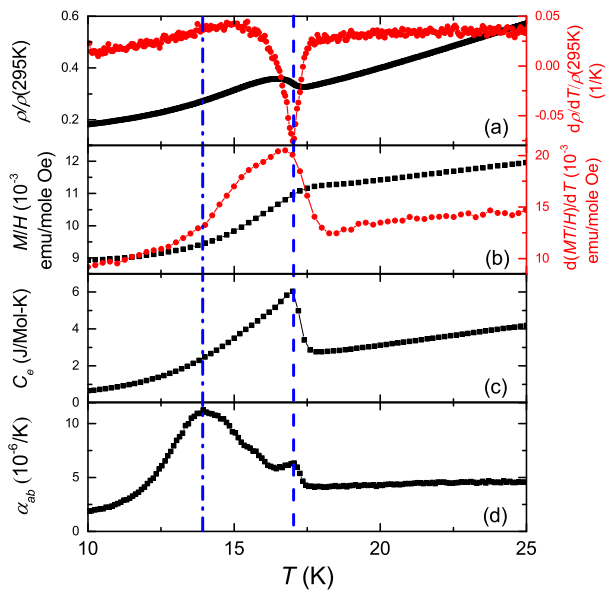}
\end{center}
\caption{(Color online) (a) Normalized electrical resistivity $\rho$/$\rho$(295 K), (b) magnetization $M/H$, (c) electronic specific heat {\itshape C}$_{e}$ and (d) thermal expansion coefficient ($\alpha_{ab}$) vs. {\itshape T} for a \UFe single crystal with {\itshape x} = 0.1.  Derivatives of $\rho$/$\rho$(295 K) and $MT/H$ with respect to {\itshape T} vs. {\itshape T} are shown in red.}  
\label{0p1}
\end{figure}

It is noteworthy that for {\itshape x} = 0.08 and 0.1, there are two separate anomalies, indicating two phase transitions. Based on the size of the jump, the feature at higher temperature is consistent with the PM-HO phase transition, while the feature at lower temperature is consistent with the transition from the HO phase into the LMAFM phase. Similar results have been reported for \UP under pressure, where the transition from the HO phase into the LMAFM phase was also observed in thermal expansion measurements.~\cite{Motoyama03,Kambe13} The sign of the thermal expansion coefficient indicates that as the sample is cooled from the HO phase into the LMAFM phase, the {\itshape ab}-plane shrinks and the {\itshape c}-axis expands, leading to an abrupt increase in the {\itshape c}/{\itshape a} ratio. An alternative scenario is that for {\itshape x} = 0.08 and 0.1, the samples are a mixture of the HO and LMAFM phase, which cannot be conclusively excluded. However, if this were the case, one would expect a smaller jump in $\alpha$ at the PM-HO phase transition, corresponding to the percentage of the HO phase. We do not see an obvious change in the size of jump for {\itshape x} = 0.08 and 0.1, indicating it is more likely due to two successive phase transitions. The reason why this phase transition from the HO to LMAFM phase is not clearly manifested in other transport or thermodynamic measurements, e.g., electrical resistivity, magnetization, or specific heat (as shown in Fig. \ref{0p1}), is not clear and requires further investigation.
%makes thermal expansion a unique bulk measurement to determine the phase boundary between HO and LMAFM phase, but also begs the question. One possible reason why this phase transition only manifests in thermal expansion is that the responses in $\alpha_{ab}$ and $\alpha_{c}$ are opposite and compensate each other. Therefore, the change in volume thermal expansion, which is related to specific heat through thermodynamic relations, becomes mild. This seems to be the case for {\itshape x} = 0.08, which can be seen in Fig.~\ref{TE}(c). However, for {\itshape x} = 0.1, even in volume thermal expansion data, this phase transition is still well observed. Further measurements are needed to inspect this phase transition.   

\section{Discussion}
\begin{figure}[!htbp]
\begin{center}
\includegraphics[angle=0,width=80mm]{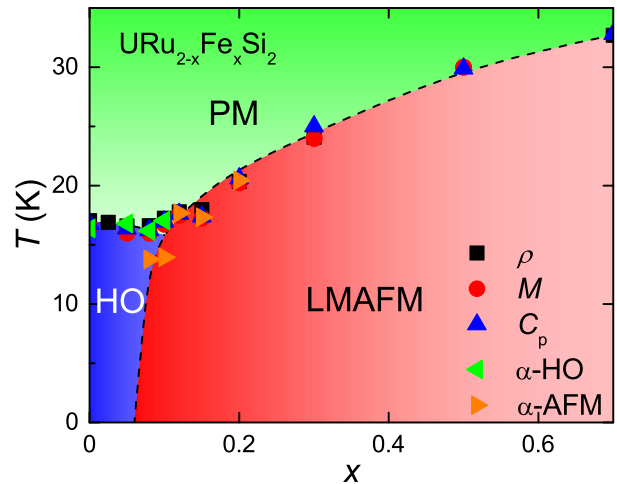}
\end{center}
\caption{(Color online) Temperature, {\itshape T}, vs. Fe concentration, {\itshape x}, phase diagram for the \UFe system. The phase diagram is based on measurements of electrical resistivity, magnetization, specific heat, and thermal expansion coefficient as a function of temperature on single crystal specimens.}
\label{PD}
\end{figure}

Based on the electrical resistivity, magnetization, specific heat and thermal expansion coefficient data, we were able to establish the temperature {\itshape T} vs Fe concentration {\itshape x} phase diagram for the \UFe single crystals shown in Fig. \ref{PD}.  In contrast to previous studies of polycrystalline samples, the PM-HO phase boundary determined in this study decreases slightly with increasing Fe substitution. The LMAFM phase becomes stable at {\itshape x} = 0.1 where the PM-LMAFM phase boundary intersects the PM-HO phase boundary. For further increases in Fe substitution, the PM-LMAFM phase boundary increases to 34~K by {\itshape x} = 0.7. 

This overall behavior is reminiscent of what was previously observed in the dependence of the HO-LMAFM phase boundary on pressure for single crystals of the parent compound URu$_{2}$Si$_{2}$, indicating that the interpretation of Fe substitution as a chemical pressure~\cite{Kanchanavatee11} still applies to the single crystals. The slight depression of the HO phase boundary with x for single crystal specimens may be caused by disorder associated with the Fe substitution, in addition to the chemical pressure. 
%as well as any discrepancies that may exist between the nominal and actual values of Fe concentration. 
It is clear that for the intermediate levels of Fe concentration, {\itshape x} = 0.8 and 0.1, there are two distinct phase transitions as observed in the thermal expansion measurements. The transitions occurring at lower temperatures help identify the phase boundary between the HO and LMAFM phases. This HO-LMAFM phase boundary can be extended to zero temperature somewhere between {\itshape x} = 0.05 and 0.08. It is now believed that the HO and LMAFM phases exhibit different symmetry with distinct order parameters. Therefore, according to Landau theory, there should be a first order phase transition between the two phases with different symmetry. However, no obvious hysteresis has been observed in thermal expansion measurements, which indicates that the HO-LMAFM transition is weakly first-order. 

\begin{figure}[!htbp]
\begin{center}
\includegraphics[angle=0,width=80mm]{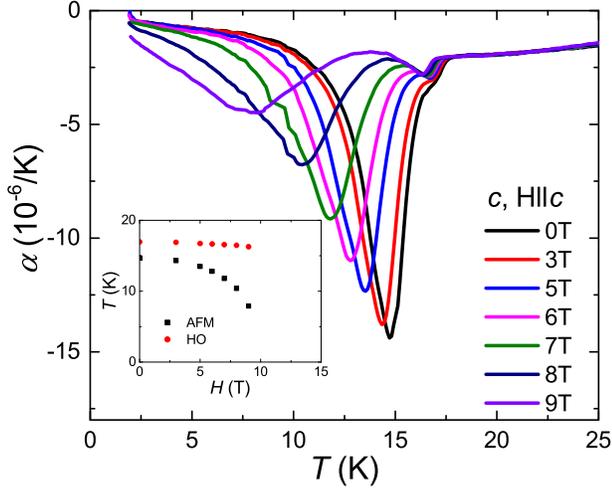}
\end{center}
\caption{(Color online) Thermal expansion coefficient along the {\itshape c}-axis vs. temperature {\itshape T} for a \UFe single crystal with {\itshape x} = 0.1 in magnetic fields up to 9~T. The magnetic field was applied along the {\itshape c}-axis. The magnetic field dependence of the two transition temperatures is shown in the inset.}
\label{Field}
\end{figure}

In order to further investigate the two distinct phase transitions detected at intermediate Fe concentrations, thermal expansion measurements were performed in magnetic fields up to 9~T on a \UFe single crystal with {\itshape x} = 0.1. The magnetic field was applied along the {\itshape c}-axis, which is the easy-axis, and the thermal expansion was measured in the same direction. The results are shown in Fig.~\ref{Field}. Both transitions are systematically suppressed to lower temperatures by the magnetic field. However, the rate of suppression is clearly different; the higher temperature transition is reduced only slightly, while the lower temperature transition is suppressed more rapidly at a rate of 7.5~K/T. The behavior of the two transition temperatures as a function of magnetic field is shown in the inset of Fig.~\ref{Field}.  These results, together with data in high magnetic field,~\cite{Aoki09,Ran16} are consistent with our hypothesis that the transition at higher temperature is the PM-HO phase transition, while the one at lower temperature is the HO-LMAFM phase transition.
%what kind of data in a field?

At the onset of the HO phase, as well as the LMAFM phase, there is a reconstruction of the Fermi surface, as revealed by transport and thermodynamic measurements.~\cite{Palstra85,Maple86,Schoenes87} For the parent compound, an energy gap associated with the HO phase, originally attributed to a charge or spin density wave,~\cite{Maple86} opens over about 40$\%$ of the Fermi surface. In order to determine how the energy gap evolves upon Fe substitution, we have evaluated the size of the gap by performing fits of relevant theoretical models to the features observed in measurements of the electrical resistivity, specific heat, and the thermal expansion coefficient that characterize the HO and LMAFM phases.

Below {\itshape T}$_{0}$, the electrical resistivity $\rho$({\itshape T}) consists of contributions from the residual resistivity, Fermi liquid electron-electron scattering, and electron-magnon scattering due to spin excitations with an energy gap $\Delta$. Since the magnons have antiferromagnetic character the following expression for $\rho$({\itshape T}) is appropriate:~\cite{Fontes99} 
\begin{multline}
  \rho(T) = \rho_{0} + AT^{2} + B\Delta^{2}(T/\Delta)^{0.5}(1+2/3(T/\Delta)  \\
   +2/15(T/\Delta)^{2})exp(−\Delta/T).
\end{multline}
The specific heat {\itshape C}$_{e}$({\itshape T}), displays a well-defined, BCS-like anomaly upon transition into the HO and LMAFM phases at {\itshape T}$_{0}$. Below the transition, the {\itshape C}$_{e}$({\itshape T}) data can be described by the expression:
\begin{equation}
  C_{e}(T) = Aexp(−\Delta/T) + \gamma T,
\end{equation}
where $\Delta$ is the gap that opens over the Fermi surface and $\gamma$ is electronic specific-heat coefficient.~\cite{Maple86} The volume thermal expansion coefficient $\beta$({\itshape T}), exhibits a BCS-like anomaly upon transition into the HO and LMAFM phases as well. Therefore, the $\beta$({\itshape T}) data can be described by a similar expression:
\begin{equation}
  \beta(T) = Cexp(−\Delta/T) + AT.
\end{equation}

\begin{figure}[!htbp]
\begin{center}
\includegraphics[angle=0,width=80mm]{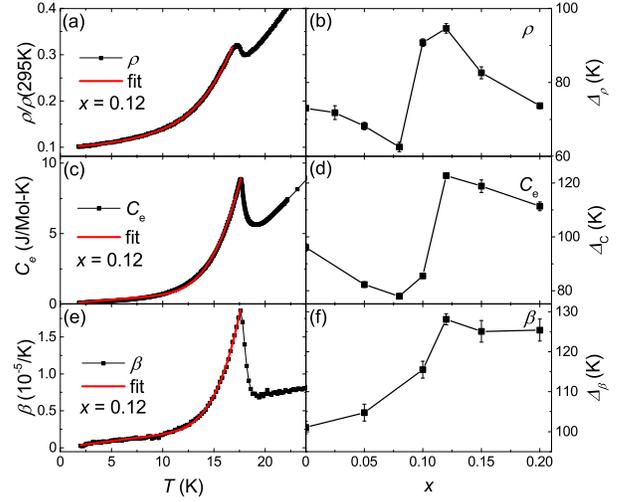}
\end{center}
\caption{(Color online) Representative fits of the expressions given in the text to the (a) electrical resistivity $\rho$({\itshape T}), (c) electronic specific heat {\itshape C}$_{e}$({\itshape T}) and (e) volume thermal expansion coefficient $\beta$({\itshape T}) data, and values of the energy gaps $\Delta_{\rho}$, $\Delta_{C}$, and $\Delta_{\beta}$, extracted from the fits (b, d, f).}
\label{Gap}
\end{figure}

The gap values extracted from the three different types of measurements are shown in Fig. \ref{Gap}, together with representative fitting curves. Although the gap values are different in magnitude, the three types of measurements show a consistent trend. The gap values for the HO phase are relatively small, $\sim$~100~K from thermal expansion, $\sim$~80~K from specific heat, and $\sim$~60~K from resistance. When crossing from the HO phase to the LMAFM phase ({\itshape x} = 0.1 and 0.12), the gap value is suddenly enhanced by $\sim$~20~K. This is consistent with what has been seen for \UP under pressure.~\cite{Jeffries07,Motoyama08} 

\begin{figure}[!htbp]
\begin{center}
\includegraphics[angle=0,width=80mm]{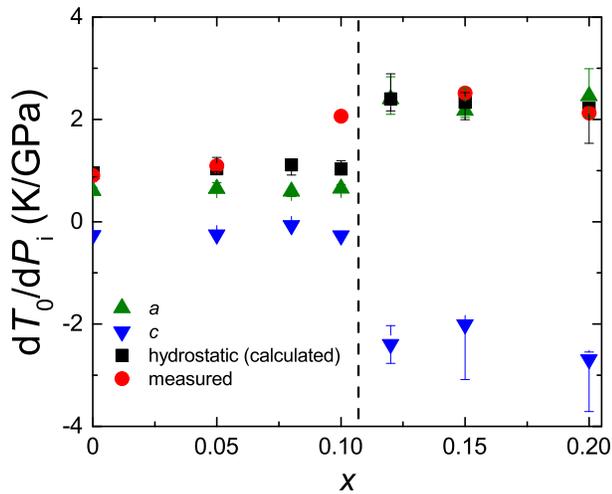}
\end{center}
\caption{(Color online) Uniaxial pressure derivatives of the transition temperature, {\itshape T}$_{0}$, estimated using the Ehrenfest relation, as well as from direct measurement.}
\label{dT0}
\end{figure}

For second-order phase transitions, the uniaxial pressure derivatives of the transition temperature, {\itshape dT}$_{0}$/{\itshape dP}, can be estimated using the Ehrenfest relation,~\cite{Barron99} 
  \begin{equation}
  dT_{0}/dP_{i} = V_{m}T_{c}\Delta \alpha_{i}/\Delta C_{p},
  \end{equation}
where {\itshape V}$_{m}$ is the molar volume which can be calculated from lattice parameters, $\Delta\alpha_{i}$ is the change in the linear ({\itshape i} = {\itshape a}, {\itshape c}) or volume ($\alpha_{v}$ = $\beta$) thermal expansion coefficient at the phase transition, and $\Delta${\itshape C}$_{p}$ is the change in the specific heat at the phase transition. The inferred uniaxial pressure derivatives are shown in Fig. \ref{dT0}. Note that for {\itshape x} = 0.08 and 0.1, both the PM-HO and PM-LMAFM phase transitions were detected in thermal expansion measurements, whereas only the PM-HO phase transition was detected in specific heat measurements. Therefore, for {\itshape x} = 0.08 and 0.1, we are only able to estimate the pressure-dependence of the transition temperature for the HO phase. The uniaxial pressure derivatives of the transition temperature have different signs for the two crystallographic orientations. Uniaxial pressure apparently applied along the {\itshape a}- or {\itshape b}-axes should produce an increase in {\itshape T}$_{0}$, whereas uniaxial pressure applied along the {\itshape c}-axis should result in a decrease in {\itshape T}$_{0}$. Another striking feature is that the pressure derivatives for the transition temperature change dramatically when crossing the HO-LMAFM phase boundary. For the HO phase, d{\itshape T$_{0}$}/d{\itshape P$_{i}$} is relatively small, 0.6~K/GPa for in-plane uniaxial pressure and 0.2~K/GPa for {\itshape c}-axis uniaxial pressure, and does not vary much with Fe concentration. On the other hand, for the LMAFM phase, d{\itshape T$_{0}$}/d{\itshape P$_{i}$} is significantly enhanced to above 2~K/GPa for both orientations. 

For hydrostatic pressure, the derivative of the transition temperature with respect to pressure can be calculated using the expression: 
  \begin{equation}
  dT_{0}/dP_{V} = 2dT/dP_{a} + dT/dP_{c},
  \end{equation}
which also shows dramatic changes upon crossing the phase boundary. The hydrostatic pressure derivatives of {\itshape T}$_{0}$, estimated using the Ehrenfest relation can be qualitatively compared with the recent results of direct measurements of electrical resistivity of URu$_{2-x}$Fe$_{x}$Si$_{2}$ under pressure,~\cite{Wolowiec16} which are also shown in Fig. \ref{dT0}. The two sets of data match quite well for the entire range of Fe concentration, except for {\itshape x} = 0.1, where the measured value is significantly higher than the one estimated using the Ehrenfest relation. As noted above, for {\itshape x} = 0.1, the value of d{\itshape T$_{0}$}/d{\itshape P} is estimated for the HO phase only. On the other hand, under applied pressure the ground state for {\itshape x} = 0.1 changes from the HO phase to the LMAFM phase, indicated by a slight kink in the {\itshape T}$_{0}$({\itshape P}) phase line. Therefore, the measured value of d{\itshape T$_{0}$}/d{\itshape P} is closer to that of the LMAFM phase. This offers an explanation for the discrepancy between the measured and estimated value of d{\itshape T$_{0}$}/d{\itshape P} at {\itshape x} = 0.1.

%For {\itshape x} = 0.2, the measured value of {\itshape dT$_{0}$}/{\itshape dP} is lower than the estimated one. Given that {\itshape dT$_{0}$}/{\itshape dP$_{i}$}, with uniaxial pressure applied along the {\itshape c}-axis, has a large negative value, this discrepancy might be caused by the conditions in the pressure cell that deviate from ideal hydrostatic conditions. In a perfect hydrostatic environment there are equal stress contributions in each direction. However, under certain conditions, there may a biased amount of uniaxial pressure along a particular direction. In the measurements of electrical resistivity for the URu$_{2-x}$Fe$_{x}$Si$_{2}$ compounds under pressure, the external pressure is applied along the {\itshape c}-axis at room temperature in a liquid medium that becomes a solid at ~ 100 K. Therefore, it is conceivable that there might be a larger pressure component along the {\itshape c}-axis which would result in a larger contribution to the measured {\itshape dT$_{0}$}/{\itshape dP$_{i}$} value. Indeed, if we weight the calculation of {\itshape dT$_{0}$}/{\itshape dP$_{i}$} in the {\itshape c}-axis with a 10$%$ increase, to reflect the uniaxial pressure bias in the {\itshape c} direction, we could match the measured values rather well, as shown in Fig. \ref{dT0}.

\section{CONCLUDING REMARKS}
We have performed thermal expansion, electrical resistivity, magnetization, and specific heat measurements on URu$_{2-x}$Fe$_{x}$Si$_{2}$ single crystals for various values of {\itshape x} between 0 and 0.7, in both the HO and LMAFM regions of the phase diagram. Our results show that the PM-HO and PM-LMAFM phase transitions are expressed differently in the thermal expansion coefficient, indicating a different coupling of the two phases to the lattice. By means of thermal expansion measurements, we also observed a possible phase transition from the HO into the LMAFM phase for intermediate levels of Fe substitution, which has not been observed in other types of measurements. In addition, the uniaxial pressure derivatives of the transition temperature, derived from thermal expansion and specific heat data, changes dramatically when crossing from the HO to the LMAFM phase. 

\section{Acknowledgement}
Single crystal growth and characterization at UCSD was supported by the US Department of Energy, Office of Basic Energy Sciences, Division of Materials Sciences and Engineering, under Grant No.~DEFG02-04-ER46105.  Low temperature measurements at UCSD were sponsored by the National Science Foundation under Grant No.~DMR 1206553.  High pressure research at UCSD was supported by the National Nuclear Security Administration under the Stewardship Science Academic Alliance Program through DOE under Grant No.~DE-NA0001841.

\end{document}